\documentclass[iopart]{jpsj3}
\usepackage{txfonts}

\title{A Field-directional Specific Heat Study on the Gap Structure of
Overdoped Ba(Fe$_{1-x}$Co$_{x}$)$_{2}$As$_{2}$}

\author{Gang Mu$^{1,5}$\thanks{E-mail: mugang@sspns.phys.tohoku.ac.jp}, Jun Tang$^2$, Yoichi Tanabe$^2$, Jingtao Xu$^2$, Bin Zeng$^3$, Bing Shen$^3$, Fei Han$^3$, Hai-Hu Wen$^{3, 4}$,
Satoshi Heguri$^1$ and Katsumi Tanigaki$^{1, 2}$\thanks{E-mail:
tanigaki@sspns.phys.tohoku.ac.jp}} \inst{$^1$Department of Physics,
Graduate School of Science, Tohoku University, Sendai
980-8578, Japan \\
$^2$WPI-AIMR, Tohoku University, Sendai 980-8578, Japan \\
$^3$Institute of Physics and Beijing National Laboratory for
Condensed Matter Physics, Chinese Academy of Sciences, Beijing
100190, China \\
$^4$ National Laboratory for Solid State Microstructures, Department
of Physics, Nanjing
University, Nanjing 210093, China \\
$^5$ State Key Laboratory of Functional Materials for Informatics,
Shanghai Institute of Microsystem and Information
Technology, Chinese Academy of Sciences, Shanghai 200050, China} 

\abst{Low-temperature specific heat is measured on the overdoped
Ba(Fe$_{1-x}$Co$_{x}$)$_2$As$_2$ ($x$ = 0.13) single crystal under
magnetic fields along three different directions. A clear anisotropy
is observed on the field dependent electronic specific heat
coefficient $\gamma(H)$. The value of $\gamma(H)$ is obviously
larger with magnetic field along [001] ($c$-axis) than that within
the $ab$-plane of the crystal lattice, which cannot be attributed to
the effect by anisotropy of the upper critical field. Meanwhile, the
data show a rather small difference when the direction of the field
is rotated from [100] to [110] direction within the $ab$-plane. Our
results suggest that a considerable part of the line nodes is not
excited to contribute to the quasiparticle density of states by the
field when the field is within the $ab$-plane. The constraints on
the topology of the gap nodes are discussed based on our
observations.}

\kword{Fe-based superconductors, gap structure, specific heat,
Ba(Fe$_{1-x}$Co$_{x}$)$_{2}$As$_{2}$, line nodes}

\begin{document}
\maketitle

\section{Introduction}
Knowledge of the gap structure will supply very important
information for the understanding of the superconducting pairing
mechanism, which is a central issue in the physics of unconventional
superconductors. Up to date, gap structure of the Fe-based
superconductors seems to be more complicated than expected, which is
found to vary substantially from family to family. The consensus has
been reached on several systems, e.g. LaFePO\cite{LaFePO1,LaFePO3},
LiFeP\cite{LiFeP}, KFe$_2$As$_2$\cite{KFe2As2-1,KFe2As2-2},
BaFe$_2$(As$_{1-x}$P$_x$)$_2$\cite{Ba122-P1,Ba122-P2,DLFeng}, and so
on, that nodes exist on the gap structure. And nodeless
superconductivity seems to have been confirmed in the
K$_x$Fe$_{2-y}$Se$_2$
system\cite{KFe2Se2-1,KFe2Se2-2,KFe2Se2-3,KFe2Se2-4} on the other
hand. However, experimental results gave rather contradicting
conclusions on this issue in other systems of the Fe-based
superconductors\cite{MuCPL1,Chien,HDing1,Hashimoto,CongRen1,MuPRB1,HeatTransport-BaK,ARSH,HDing2,CLSong,penetration-depth,HeatTransport-BaCo,SH-BZeng}.
Furthermore, recent experiments reveal that the gap structure can
modulate with the doping concentration even in the same family. In
the case of electron-doped (Co- or Ni- doped) 122 system, the
increase in the gap anisotropy and even the emergence of gap nodes,
as the doping content increases in the overdoped region, are
evidenced by many experimental
methods\cite{penetration-depth,HeatTransport-BaCo,anneal4,MuPRB2,CongRen2,optical-conductivity}.
This behavior is attributed to the enhancement of intraband
interaction and pair scattering between the electron-like Fermi
surfaces as the system is doped away from the optimal
point\cite{Mazin-Review}. Many efforts have been made to obtain more
information about the location and direction of the nodes.
Directional penetration depth measurements have suggested
latitudinal circular line nodes located at the finite $k_z$ wave
vector or a point (or extended area) polar
node\cite{penetration-depth}. Based on the directional heat
transport measurements, a simplified two-band model of the Fermi
surface (FS) was proposed, where line nodes on one FS sheet with
strong 3D character and a deep minimum of the gap on the other FS
sheet with quasi-2D character are
suggested\cite{HeatTransport-BaCo}. A theoretical analysis has shown
that the line nodes may form vertical loops in the present of
hybridization between two electron pockets.\cite{loop-nodes} For the
BaFe$_2$(As$_{0.7}$P$_{0.3}$)$_2$ system, angle-resolved thermal
conductivity measurements have suggested the closed nodal loops at
the flat parts of the electron FS\cite{Ba122-P2}. In contrast, an
angle resolved photoemission spectroscopy (ARPES) study has revealed
a horizontal circular line node on the hole FS around the Z point at
the Brillouin zone boundary\cite{DLFeng}.

Specific heat (SH) is a powerful bulk probe for investigating the
gap structure of the unconventional superconductors. The variation
of electronic SH ($C_{el}$) versus temperature ($T$) strongly
depends on the gap structures of the superconducting
materials\cite{review1,review2}. In our previous
studies\cite{MuPRB2}, we have observed a clear $T^2$ term in
$C_{el}$ of the overdoped Ba(Fe$_{1-x}$Co$_{x}$)$_{2}$As$_{2}$,
giving a direct evidence for the presence of line nodes in the
energy gap, in sharp contrast to that observed in the underdoped and
optimal doped samples which show a rather small anisotropy in the
energy gap.  In the mixed state, the applied field ($H$) can induce
vortices with a supercurrent flowing perpendicularly to $H$. The
low-energy quasiparticles will undergo a Doppler shift induced by
the supercurrent, which can give rise to a considerable enhancement
of $C_{el}$ for a nodal superconductor\cite{Vol1,Vol2}. This is
called the Volovik effect. Doppler-shift energy is determined by
$\delta E \propto \vec{v}_{s}\cdot \vec{v}_{F}$, where $\vec{v}_{s}$
and $\vec{v}_{F}$ are the supercurrent velocity and Fermi velocity
respectively. We note that $v_F$ is proportional to $\nabla_k E(k)$
and thus it is perpendicular to the Fermi surface. Moreover, Volovik
effect works mainly at the places near the nodes. So the Volovik
effect is the strongest (weakest) when $H$ is parallel
(perpendicular) to the Fermi surface where the nodes reside. This
provides valuable information on the topology of the line nodes,
when we compare the SH data obtained with $H$ along different
directions of the crystal lattice.

In order to obtain further information on the topology of the line
nodes, here we report a field-directional specific heat study on the
overdoped Ba(Fe$_{1-x}$Co$_{x}$)$_2$As$_2$. The unambiguous
anisotropic behaviors related to the Volovik effect imposed on line
nodes are confirmed by the fact that the electronic SH coefficient
$\gamma(H)$ increases more quickly when the field is parallel to the
$c$-axis than that within the $ab$-plane. The possible constraints
on the topology of line nodes are discussed based on our
observations.

\begin{figure}
\begin{center}
\includegraphics[width=0.6\textwidth]{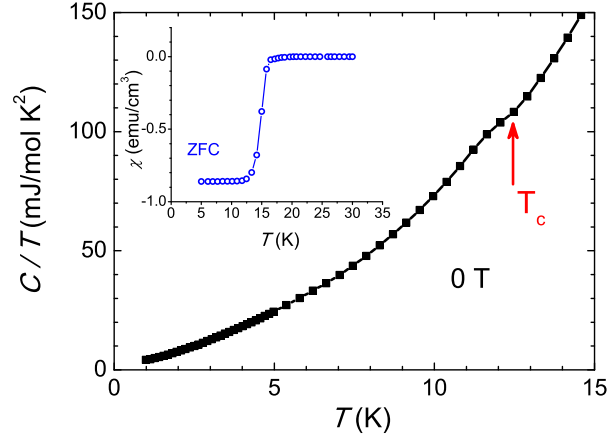}
\end{center}
\caption {(Color online) Main frame: temperature dependence of SH
coefficient ($C/T$) for the overdoped
Ba(Fe$_{1-x}$Co$_{x}$)$_{2}$As$_{2}$ with $x$ = 0.13 under zero
field. The red arrow denotes the specific heat anomaly due to the
superconducting transition. Inset: dc susceptibility for the same
sample measured using the zero-field-cooling process. The dc field
of 10 Oe was applied perpendicular to the $c$-axis. } \label{fig1}
\end{figure}

\section{Experimental Details and Sample Characterization}

The Ba(Fe$_{1-x}$Co$_{x}$)$_{2}$As$_{2}$ single crystals were grown
by a self-flux method using FeAs as the flux. As-grown samples were
annealed under high vacuum at 800 $^o$C for 20 days. An identical
crystal is used for all the measurements in this paper, which has a
dimensions of about 2.4$\times$2.0$\times$0.2 mm$^3$ and the mass of
about 6.5 mg. The dc magnetization measurements were made with a
superconducting quantum interference device (Quantum Design, MPMS).
Specific heat was measured with magnetic field along three
directions (see figure~2(a)) in the field-cooling mode. The data
obtained with $H$ within $ab$-plane were collected with a Helium-3
system attached with the physical property measurement system
(Quantum Design, PPMS), while those with $H$ along $c$-axis were
obtained from PPMS. We employed the thermal relaxation technique to
perform the specific heat measurements. The thermometer has been
calibrated under different magnetic fields beforehand.

The superconducting transition of the selected sample with nominal
doping contents $x$ = 0.13 is checked by the dc magnetization and
specific heat measurements. The present sample was determined to be
in the overdoped region of the phase diagram\cite{MuPRB2}. It has
been confirmed by many measurements that no magnetic order exists in
this doping region\cite{phase-diagram1,phase-diagram2}. So it
supplies a very clean platform to study the behaviors of specific
heat. In figure~1, we show the temperature dependence of the SH
coefficient $C/T$ up to 15 K under zero field. The kink at about
12.5 K indicated by red arrow is the superconducting transition.
This temperature corresponds to the end point of the diamagnetic
transition, as shown in the inset of figure~1. The superconducting
volume fraction estimated from the magnitude of the dc
susceptibility suggests a nearly full Meissner fraction for the
present sample.

\section{Results}

\begin{figure}
\begin{center}
\includegraphics[width=13cm]{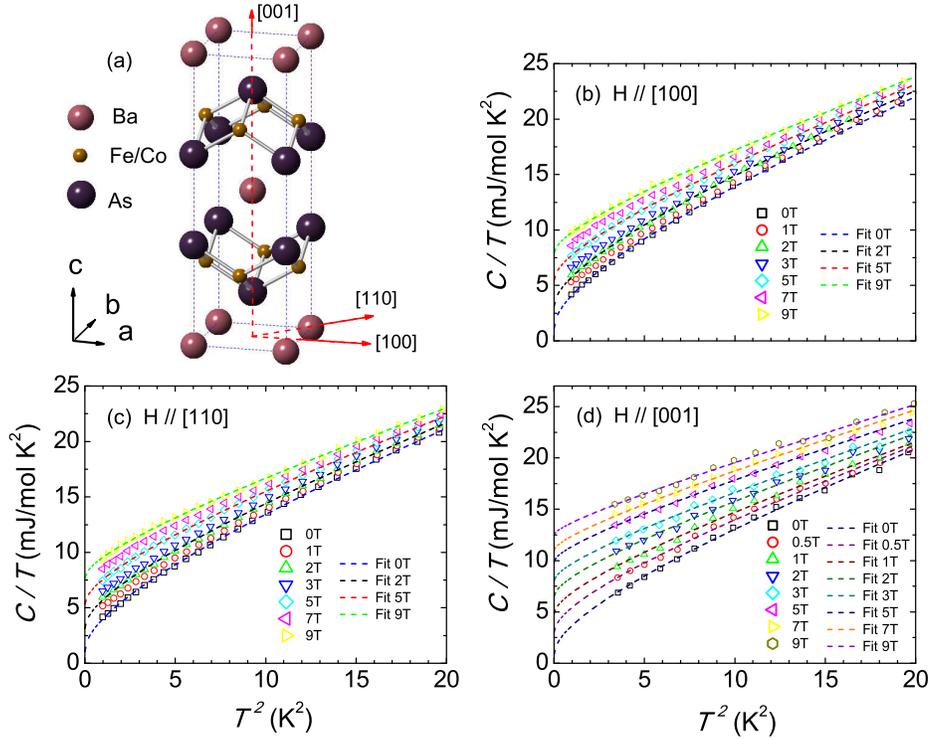}
\end{center}
\caption {(Color online) (a) Crystal structure of
Ba(Fe$_{1-x}$Co$_{x}$)$_{2}$As$_{2}$. Red arrowed lines represent
the three directions of the crystal lattice along which the magnetic
field is applied. (b)-(d) show the raw data of the SH for the sample
with x = 0.13 under fields aligned with the [100], [110], and [001]
directions of the crystal lattice, respectively. The dashed lines
represent the results of theoretical fitting (see text). }
\label{fig2}
\end{figure}

We measured SH of Ba(Fe$_{1-x}$Co$_{x}$)$_{2}$As$_{2}$ with $x$ =
0.13 under field along the [100], [110], and [001] directions of the
crystal lattice. A schematic representation of the three directions
is given in figure~2(a). The raw data of SH under $H$ of these three
directions are plotted as $C/T$ vs $T^2$ in figures~2(b)-(d),
respectively. Here we focus on the behaviors of our data in the
low-$T$ region below 4.5 K. One can see that the behaviors of the
data with $H$ along [100] and [110] directions are quite similar
with each other. While the data with $H$ along the [001] direction
increase more quickly with $H$. Nevertheless, all the three sets of
data show clear negative curvatures in the temperature range we
studied. In our previous work\cite{MuPRB2}, we have attributed this
behavior to the presence of the $T^2$ term in the electronic SH,
which is consistent with the prediction for the superconductors with
line nodes in the energy gap. No Schottky anomaly can be seen in our
data suggesting a very low concentration of the magnetic impurity in
our sample. The data were then fitted by the following equation
\begin{equation}
C(T,H) = \gamma(H)T+\alpha(H) T^2 + \beta T^3,\label{eq:1}
\end{equation}
where $\gamma(H)$ is the electronic SH coefficient under $H$,
$\alpha(H)$ is the coefficient of the $T^2$ term under $H$, and
$\beta$ is the phonon SH coefficient. The value of $\beta$ was found
to be almost independent of the field, with deviation below the
scale of 5\%. Consequently, here we average the value of $\beta$
under different fields and fix it when fitting the data using
equation~(1). The effect of small fluctuations of $\beta$ is
transferred to the error bar of the resulting fitting parameters
$\gamma(H)$ and $\alpha(H)$ (see figure~3). The fitting results are
displayed by the dashed lines in figure~2. Only four selected
fitting curves are shown in figures~2(b) and (c) respectively for
clarity. It is clear that these curves describe the
negative-curvature features commendably. From the fitting we find
that the residual electronic SH coefficient $\gamma (H=0)$ for the
three sets of data shows very close value of about 1 mJ/mol K$^2$.
This value is smaller than the previously reported results in the
similar systems\cite{anneal4,SH-direcitional1}, suggesting the high
quality of our sample.

We note here multi-gap effect should be considered when fitting our
data in principle in such a multi-band system, because the magnitude
and structure of the gaps on different FS sheets may be rather
different. However, one difficulty is that the number of the
parameters will be too many to give a reliable fitting, if we take
all the five bands into account simultaneously. Fortunately, the
electronic SH is mainly contributed from the quasiparticles with
heavy mass, which have been found to come from hole pockets by band
calculations. So our observations imply that line nodes may exist on
one of the hole pockets. Moreover, electronic SH from the pockets
with line nodes will overcome that from fully gapped pockets in the
low temperature limit, because the latter should follow an
exponential law with temperature. Consequently, it is safe to
neglect the contributions from other fully gapped pockets when
fitting the low-temperature data. For the same reason, we will only
consider the FS sheet with line nodes when discussing the possible
topologies of the gap nodes in the next section.

The fitting parameters $\gamma(H)$ and $\alpha(H)$ are shown in
figures~3(a) and (b). The $x$-coordinate is normalized with the
upper critical field $H_{c2}$, so as to eliminate the effect of
anisotropy of $H_{c2}$ on the $H$ dependent data. As we have
stated\cite{MuPRB2}, $H_{c2}$ within the $ab$-plane is about 28 T,
from which the $H_{c2}$ value along the $c$-axis can be estimated
using the its anisotropy ($\Gamma\equiv H^{ab}_{c2}/H^c_{c2}$). From
the reported high-$H$ experiments, we know that $\Gamma$ decreases
when reducing $T$ and finally reaches about 1.1 at 0.7 K for the
near-optimal doped
Ba(Fe$_{1-x}$Co$_{x}$)$_2$As$_2$\cite{Hc2-1,Hc2-2}. Here we take the
value $\Gamma\sim 1.3$ for the present overdoped case, which would
be no less than the actual value. Such a treatment will not affect
the conclusions described below, because a smaller $\Gamma$ would
result in the enhancement of the anisotropic features.

\begin{figure}
\begin{center}
\includegraphics[width=7cm]{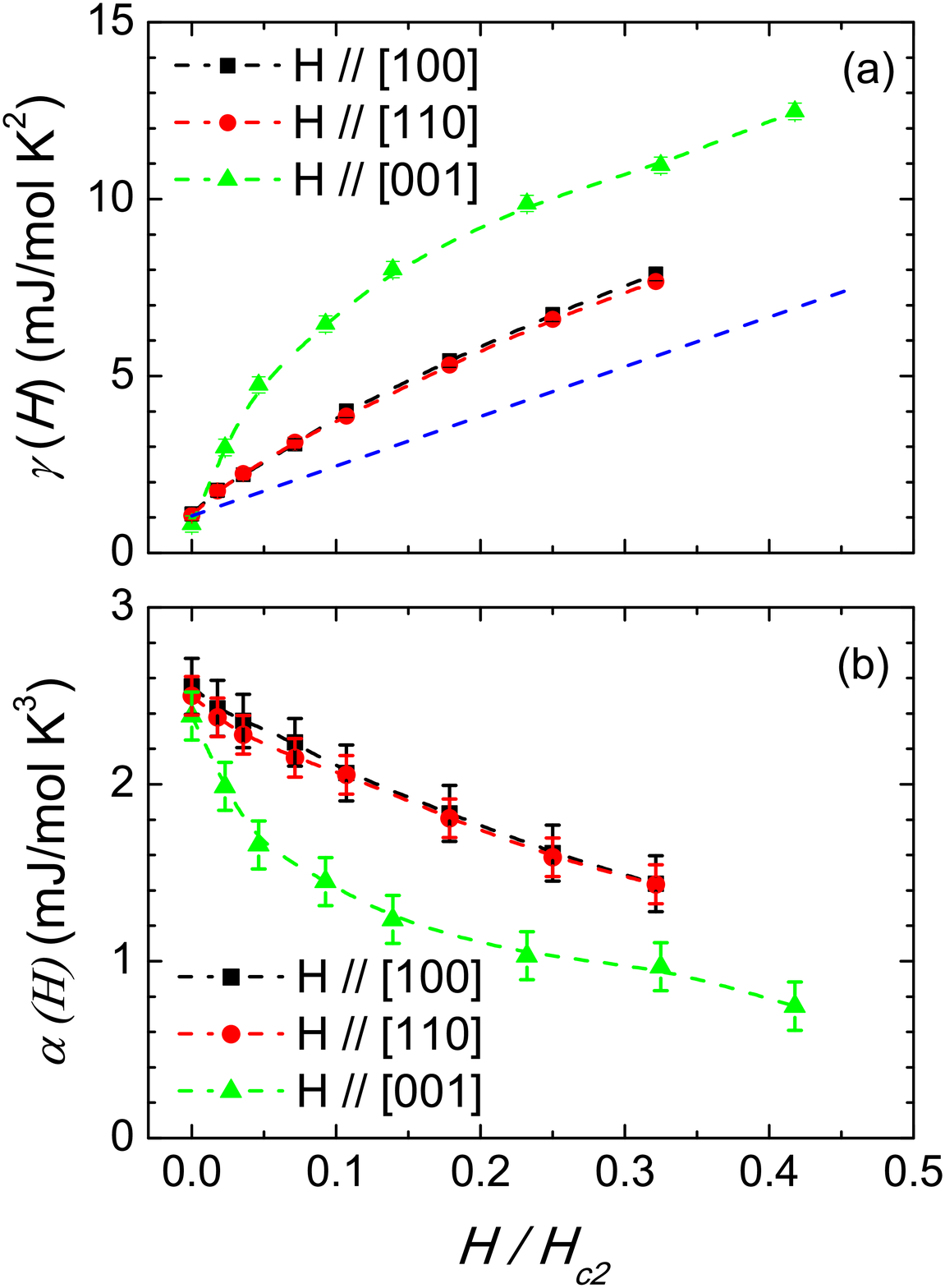}
\end{center}
\caption {(Color online) Field dependence of $\gamma(H)$ and
$\alpha(H)$ with $H$ along three directions. $H$ is normalized with
the upper critical field $H_{c2}$. The blue dashed line in (a) shows
the theoretical curve for the case with an isotropic gap. Other
dashed lines are guides to the eye.} \label{fig3}
\end{figure}

Both $\gamma(H)$ and $\alpha(H)$ show small deviations among the
three directions when $H$ is approaching zero, which confirms the
reliability of our data even though the data with $H$ along $c$-axis
is only measured down to 1.9 K. From figure~3(a) one can see that
all the three sets of data are clearly above the blue dashed line,
which represents the theoretical curve for the case with an
isotropic gap. This is consistent with the features induced by the
Volovik effect in a nodal superconductor. It is clear that
$\gamma(H)$ with $H$ aligned with $c$-axis (hereafter abbreviated as
$\gamma^{c}(H)$) increases more quickly than that within the
$ab$-plane (abbreviated as $\gamma^{ab}(H)$). We found that
$\gamma^{ab}(H)$ takes up about 70\% of $\gamma^{c}(H)$ when the
reduced $H/H_{c2}$ equals to 0.3. In sharp contrast, the data within
the $ab$-plane remain almost unchanged within the extent of error
bar when $H$ is rotated from [100] to [110] direction. Considering
the fact that the increase of $\gamma(H)$ is induced by Volovik
effect, our present observations suggest that the Volovik effect is
stronger when $H$ is applied along the $c$-axis than that within the
$ab$-plane. The anisotropic features of $H$ dependent $\alpha(H)$
shown in figure~3(b) also support such an argument because the
decrease of $\alpha(H)$ is associated with the Volovik effect. The
$T^2$ term reflects the V-shape of the density of states (DOS) at
the nodes\cite{review2}. And the Volovik effect induced by magnetic
field will destroy the V-shape of DOS in the low-energy limit. We
note that our argument here is different from previously reported
field-directional SH data on similar systems, where the anisotropic
behaviors of $\gamma(H)$ were only attributed to the effect of
anisotropy of $H_{c2}$\cite{SH-direcitional1} which would result in
an overestimation of $\Gamma$ ($>$ 2). Obviously, the findings here
will provide a hint for investigating the topology of the nodes,
which will be discussed in the next section.

\section{Discussion}

As we have stated, different topologies of the nodes or gap minima
have been proposed from different measurements and calculations.
Generally speaking, three typical models can be anticipated, namely
vertical line nodes, horizontal line nodes and vertical loop nodes.
Also the significant influence by the three-dimensional (3D)
features of the Fermi surface (FS) has been noticed by
band-calculations\cite{Graser1,calculation-122P}. Consequently, here
we consider the constraints by our data on the line nodes based on
the two simple models, where vertical and horizontal line nodes are
located on one of the FSs with a certain extent of 3D character. The
situation of vertical loop nodes can be considered as a mixture of
the case of the simple vertical and horizontal line nodes, in
principle. As shown in figure~4, the longitudinal section of the FS
sheet is shown by two black lines. The red line in figure~4(a) and
red circle in figure~4(b) represent the positions of vertical and
horizontal line nodes, respectively. The anisotropy of Fermi
velocity and other details of the Fermi surfaces are not taken into
account in the present simple models. Considering the fact that the
hole pockets have heavier mass quasiparticles than the electron
pockets, most of the electronic SH is contributed from the hole FS
sheets. So the FS sheets shown in figure~4 should most likely be one
of the hole FS sheets around the $\Gamma$ point.

We first check the situation of the vertical-node case as shown in
figure~4(a). When $H$ is applied parallel to $k_z$, which is shown
by the blue arrow $H_1$, the segments of line nodes on the FS with
no or small 3D features will experience a strong Volovik effect,
while that with clear 3D features only undergo a depressed Volovik
effect induced by the projection of $H$ on the tangent surface
($H_{\parallel}=H_1 \cos \theta$), where angle $\theta$ describes
the deviation of the FS from the direction of $k_z$. When $H$ is
perpendicular to $k_z$, the situation will be rather complex because
the direction can be rotated by 360$^o$. The Volovik effect is the
weakest if $H$ is rotated to the nodal direction (see $H_2$ in
figure~4(a)) because only the segments of nodes on the 3D-dispersed
FS can experience the Volovik effect with a projected field
$H_{\parallel}=H_2 \sin \theta$. The situations with $H$ along other
directions may vary depending on the number of the vertical line
nodes and their distribution on the FS. Nevertheless, the clear
anisotropy between $\gamma^{c}(H)$ and $\gamma^{ab}(H)$ requires
that the angle $\theta$ or the proportion of the 3D-dispersed FS
should not be too large.

\begin{figure}
\begin{center}
\includegraphics[width=9cm]{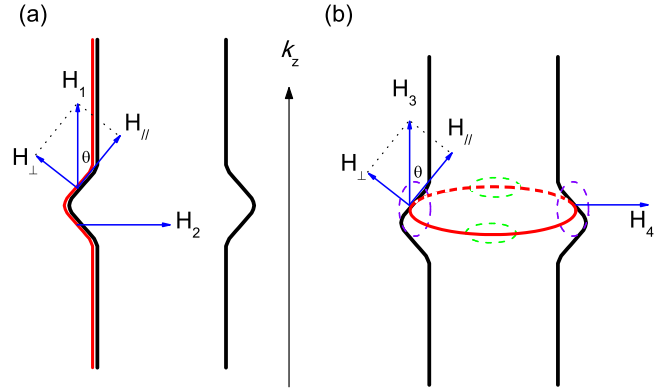}
\end{center} \caption {(Color online) Schematic diagram of two
typical models of line nodes on one of the Fermi surfaces with a 3D
character. Here we only display a longitudinal section of the Fermi
surface shown by the two black lines. The vertical (a) and
horizontal (b) line nodes are shown by the red lines.} \label{fig4}
\end{figure}

One problem in the scenario described above is that $\gamma^{ab}(H)$
should show some variation in principle when $H$ is rotated within
$ab$-plane, which was not observed in our data. One possibility is
that the number of nodes is large (e.g. larger than 4). In this
case, the angle dependent behavior of $\gamma^{ab}(H)$ will be
unconspicuous and difficult to be detected from the measurements.
Another explanation may be given by calculations based on an
extended-s-wave case\cite{Graser2}, where an attenuation of the SH
vibration is predicted because of the elliptical FS pockets near the
M points. It seems that both cases do not support the $d$-wave
symmetry of the energy gap. At present we can't rule out the
possibility that both [100] and [110] directions deviate from the
nodal direction, and consequently we failed to observe the
difference between the two directions. This may need further
clarification by detailed angle-resolved SH measurements.

The situation becomes somewhat different if we have the horizontal
line nodes, as shown in figure~4(b). When $H$ is parallel to $k_z$
(see $H_3$), the whole nodal line will experience a depressed
Volovik effect induced by $H_{\parallel}= H_3 \cos \theta$. Whereas
when $H$ is applied perpendicular to $k_z$ (see $H_4$), the segments
of nodes on the areas of the FS marked by the green circles will
experience a strong Volovik effect because $H$ is roughly parallel
to these parts of FS. Meanwhile, the segments of nodes marked by the
violet circles will experience a depressed Volovik effect induced by
$H_{\parallel}= H_4 \sin \theta$. The fact that $\gamma^{c}(H)$ is
clearly larger than $\gamma^{ab}(H)$ in our data means that the
overall Volovik effect when $H$ is parallel to $k_z$ should exceed
that with $H$ perpendicular to $k_z$. This implies that the angle
$\theta$ cannot be too large. The advantage of this model is that
the unobservable vibration of $\gamma^{ab}(H)$ can be explained
naturally. We note that this horizontal-line-nodes model is rather
similar to that observed in BaFe$_2$(As$_{0.7}$P$_{0.3}$)$_2$ by the
ARPES measuremens\cite{DLFeng}.

\section{Concluding remarks}

In summary, we studied the low-temperature SH on the overdoped
Ba(Fe$_{1-x}$Co$_{x}$)$_{2}$As$_{2}$ with magnetic field along three
different directions. Clear anisotropic behaviors are observed from
$H$ dependent data. The electronic SH coefficient $\gamma(H)$
increases more quickly when $H$ is in the $c$-axis direction than
that within the $ab$-plane, whereas the data remain unchanged within
our resolution when $H$ is rotated within the $ab$-plane. Our
results suggest that a considerable portion of the line nodes is not
excited completely to contribute to the density of states when $H$
is in the $ab$ in-plane. These conclusions supply important
constraints when investigating the topologies of the line nodes in
this system.

\begin{acknowledgment}

We acknowledge discussions with Dr. Yue Wang. The research is
partially supported by Scientific Research on Priority Areas of New
Materials Science Using Regulated Nano Spaces, the Ministry of
Education, Science, Sports and Culture, Grant-in-Aid for Science,
and Technology of Japan. The work is partially supported by Tohoku
GCOE Program and by the approval of the Japan Synchrotron Radiation
Research Institute (JASRI). This work is partially supported by the
Knowledge Innovation Project of Chinese Academy of Sciences (No.
KJCX2-EW-W11). G M expresses special thanks to Grants-in-Aid for
Scientific Research from the Japan Society for the Promotion of
Science (JSPS) (Grant No. P10026).

\end{acknowledgment}

\end{document}